\documentclass[12pt]{article}

\usepackage{color}

\catcode`ð=\active
 \defð{\u{g}}
 \catcode`Ð=\active
 \defÐ{\u{G}}
 \catcode`Ý=\active
 \defÝ{\. I}
 \catcode`ö=\active
 \defö{\"{o}}
 \catcode`Ö=\active
 \defÖ{\"O}
 \catcode`ü=\active
 \defü{\"{u}}
 \catcode`Ü=\active
 \defÜ{\"{U}}
 \catcode`Þ=\active
 \defÞ{\c{S}}
 \catcode`þ=\active
 \defþ{\c{s}}
 \catcode`ý=\active
 \defý{{\i}}
 \catcode`ç=\active
 \defç{\d{c}}
 \catcode`Ç=\active
 \defÇ{\d{C}}

\begin{document}
\title{\textbf{Analytical Solution of two-body spinless Salpeter Equation for Hellmann Potential}}

\author{{\small Altug Arda\footnote{arda@hacettepe.edu.tr}}\\
{\small \emph{Department of Physics Education, Hacettepe University}}, \\
{\small \emph{06800,
Ankara, Turkey}}}
\date{}

\maketitle

\begin{abstract}
Approximate bound state solutions of the spinless Salpeter equation for the Hellmann potential are studied for heavy particles. By using functional analysis method, an analytical expression for the energy levels, and the corresponding eigenfunctions of the system are obtained in terms of the hypergeometric functions. The analytical results for the Yukawa and Coulomb potentials are also studied as special cases.


\end{abstract}
\newpage

\section{Introduction}
 The covariant Bethe-Salpeter equation having a formal connection to the
relativistic quatum field theory is suitable for describing bound states
of pair of two particles. [1, 2]. The Salpeter equation is obtained by eliminating the time-like dependence variable by considering a static or instantaneous interactions [3, 4]. It can be considered as an approximation to the Bethe-Salpeter formalism by neglecting the spin degree of freedom, and restricting to positive energy solutions [3-5]. On the other hand, the Salpeter equation represents simply semi-relativistic generalization of the Schrödinger equation including the exact free-particle relativistic relation between energy and momentum [6]. This equation is used when relativistic kinetic term effect is not negligible. So, it is suitable for bosons as well as the spin averaged spectra of bound states of fermions [3, 4, 7]. In addition, the two-body spinless Salpeter equation gives acceptable results, especially, for the bound states of deuteron, exciton and mesons [6, 8]. The solution of one-body and/or two-body spinless Salpeter equation [8] has some difficulties because of the form of the kinetic terms having nonlocal operators in the configuration-space representation. That is why, the solutions of this equation have been studied only by few authors, especially by Lucha and co-workers [9-23]. It seems that these some papers about analytical bound state solutions for the spinless Salpeter equation in literature for different potentials such as Coulomb-type potential [3-5], kink-like potential [23], Hulthen potential [24] and power-law potential [25]. In addition, Ikhdair and co-workers have studied the mass spectra of heavy quarkonia by solving the spinless Salpeter equation [26-28]. In the present work, we deal with the exact analytical solutions of the two-body Salpeter equation for the Hellmann potential [29] which has many applications in different areas such as electron-core and electron-ion interactions, inner-shell ionization problems and solid-state physics [30].

The organization of this work is as follows. In Section II, we present the required expressions within the spinless Salpeter formalism. In Section III, we study the analytical solutions for the bound states and the corresponding wavefunctions for the Hellmann potential by using functional analysis method. We also present the analytical results for the Yukawa and Coulomb potentials, separately. In Section IV, we give our conclusions.

\section{S\lowercase{pinless} S\lowercase{alpeter} E\lowercase{quation}}

As stated in Ref. [31], the approximate Bethe-Salpeter equation makes it possible to obtain the spectrum of two bound, interacting quark-antiquark of masses $m_1$ and $m_2$. In Ref. [32], the authors have presented analytical solutions to the spinless $s$-wave Salpeter equation for two fermions interacting via a spherical potential.

The spinless Salpeter equation for a spherical potential takes the form [31-33]
\begin{eqnarray}
\left[\sqrt{-\Delta+m^{2}_{1}\,}+\sqrt{-\Delta+m^{2}_{2}\,}+V(r)-M\right]\psi(r)=0\,,
\end{eqnarray}
where $\Delta=\nabla^2$ and $M$ is the mass of the bound state. The total wavefunction has the form $\psi(r)=R_{n\ell}\,(r)Y_{\ell\,m}(\theta,\phi)$ where $R_{n\ell}\,(r)$ represents the radial wave function while $Y_{\ell\,m}(\theta,\phi)$ represents the angular dependence of the wave functions with quantum numbers ($n, \ell, m$).

The approximation
\begin{eqnarray}
\sqrt{-\Delta+m^{2}_{1}\,}+\sqrt{-\Delta+m^{2}_{2}\,}=m_1+m_2-\frac{\Delta}{2\mu}-\frac{\Delta^2}{8\eta^{3}}-\ldots\,,
\end{eqnarray}
with
\begin{eqnarray}
\mu=\frac{m_{1}m_{2}}{m_1+m_2}\,\,;\eta=\mu\,\left(\frac{m_{1}m_{2}}{m_{1}m_{2}-3\mu^2}\right)\,.
\end{eqnarray}
can be used if one considers the case of heavy interacting particles [6]. From the Hamiltonian containing the relativistic corrections up to order $(\upsilon^2/c^2)$ can be obtained the following Schrödinger-like equation [31]
\begin{eqnarray}
\left[-\frac{\hbar^2}{2\mu}\frac{d^2}{dr^2}+\frac{\ell(\ell+1)\hbar^2}{2\mu r^2}+W_{n\ell}\,(r)-\frac{W^{2}_{n\ell}\,(r)}{2\tilde{m}}\right]R_{n\ell}\,(r)=0\,,
\end{eqnarray}
where
\begin{eqnarray}
W_{n\ell}\,(r)=V(r)-E_{n\ell}\,,\nonumber\\
\tilde{m}=\frac{\eta^3}{\mu^2}=\frac{m_{1}m_{2}\mu}{m_{1}m_{2}-3\mu^2}\,,
\end{eqnarray}
with the total energy of the system $E_{n\ell}=M-m_{1}-m_{2}$.

We solve approximately Eq. (4) for the Hellmann potential below with the help of the functional analysis method.

\section{B\lowercase{ound} S\lowercase{tate} S\lowercase{olutions}}

The Hellmann potential is the sum of the Yukawa and Coulomb potentials having the form
\begin{eqnarray}
V(r)=-\frac{A}{r}+B\,\frac{e^{-\,Cr}}{r}\,,
\end{eqnarray}
where $A$ and $B$ are the strengths of potentials, respectively, $C$ is the screening parameter. The parameters $A$ and $B$ could be positive or negative while it is assumed $C$ is positive [30].

Inserting the above potential into Eq. (4), and using the following approximation which gives consistent results for small values of parameter $C$ [34]
\begin{eqnarray}
\frac{1}{r^2} \simeq \frac{C^2}{(1-e^{-\,Cr})^2}\,,
\end{eqnarray}
we have
\begin{eqnarray}
\left\{\frac{d^2}{dr^2}-\frac{C^2\ell(\ell+1)}{(1-e^{-\,Cr})^2}+\frac{M_2}{M_1}\,\frac{C^2}{(1-e^{-\,Cr})^2}\left(A^2+B^2e^{-2Cr}-2\frac{AB}{C}\,e^{-\,Cr}\right)\nonumber\right.\\-\left.
2M_2\left(1+\frac{E}{M_1}\right)\,\frac{C}{1-e^{-\,Cr}}\left(-A+Be^{-\,Cr}\right)+2M_{2}E\left(1+\frac{E}{2M_1}\right)\right\}R_{n\ell}\,(r)=0\,,
\end{eqnarray}
where we set $M_1=\tilde{m}$ and $M_2=\mu$.

By defining a new variable $z=(1-e^{-\,Cr})^{-1}$, writing the trial wave function as $R_{n\ell}\,(r)=z^{\alpha}(1-z)^{\beta}\omega(z)$, Eq. (8) takes the form
\begin{eqnarray}
&&z(1-z)\frac{d^2\omega(z)}{dz^2}+[1+2\alpha-2(\alpha+\beta+1)]\frac{d\omega(z)}{dz}\nonumber\\&-&\left[(\alpha+\beta)(\alpha+\beta+1)-\ell(\ell+1)
+\frac{M_2}{M_1}(A^2+B^2)+2\frac{M_2}{M_1}\frac{AB}{C}\right]\omega(z)=0\,.
\end{eqnarray}
If we set the parameters $\alpha$ and $\beta$ as
\begin{eqnarray}
\alpha^2&=&-2\frac{M_2}{C}\left(1+\frac{E}{2M_1}\right)\left(B+\frac{E}{C}\right)-\frac{M_2}{M_1}B\left(B+\frac{E}{C}\right)\,,\nonumber\\
\beta^2&=&\ell(\ell+1)-\left(A+\frac{E}{C}\right)\left[\frac{M_2}{M_1}A+2\frac{M_2}{C}\left(1+\frac{E}{2M_1}\right)\right]\,.
\end{eqnarray}
Eq. (9) can be compared with the hypergeometric equation having the form [35]
\begin{eqnarray}
z(1-z)\frac{d^2\phi(z)}{dz^2}+[c-(a+b+a)z]\frac{d\phi(z)}{dz}-ab\phi(z)=0\,.
\end{eqnarray}

Thus, the general solution becomes as a sum of two linearly independent solutions
\begin{eqnarray}
\phi(z)=N'\,_{2}F_{1}(a,b;c;z)+N''z^{1-c}\,_{2}F_{1}(a+1-c,b+1-c;2-c;z)\,.
\end{eqnarray}
The hypergeometric function in the above is written as an infinite sum [35]
\begin{eqnarray}
\,_{2}F_{1}(a,b;c;z)=\sum_{k=0}^{\infty}\frac{(a)_{k}(b)_{k}}{(c)_{k}k!}z^{k}\,,
\end{eqnarray}
with $(a)_{k}=a(a+1)\ldots (a+k-1)$ Pochhammer symbol. We search the bound state solutions for $R_{n\ell}\,(r)$ in Eq. (8) which means that $R_{n\ell}\,(r)$ has to be finite for $r \rightarrow 0$ and $r \rightarrow \infty$. In order to ensure these physical requirements we set $N''=0$ in Eq. (12) for $z \rightarrow \infty$. So, the solution of Eq. (9) can be expressed in terms of the hypergeometric function (without any normalization constant)
\begin{eqnarray}
\omega(z)=\,_{2}F_{1}(a,b;c;z)\,,
\end{eqnarray}
where the parameters $a$, $b$ and $c$ are given by
\begin{eqnarray}
a&=&\frac{1}{2}\,(1+2\alpha+2\beta)-A_{I}\,,\nonumber\\
b&=&\frac{1}{2}\,(1+2\alpha+2\beta)+A_{I}\,,\nonumber\\
c&=&1+2\alpha\,.
\end{eqnarray}
where $A_{I}=\sqrt{(\ell+1/2)^2-(M_2/M_1)(A^2+B^2-2AB/C)\,}$. When either $a$ or $b$ equals to a negative integer $-n$, the solution written as an infinite sum in Eq. (13) gives a polynomial of $n$th degree in terms of $z$, which gives also the quantization condition of the system under consideration
\begin{eqnarray}
a=-n\,,\,\,\,n=0, 1, 2, \ldots\,.
\end{eqnarray}
Substituting Eq. (10) into Eq. (16) and using Eq. (15), we obtain the energy spectrum of the spinless Salpeter equation for the Hellmann potential as
\begin{eqnarray}
E=\pm\frac{1}{4M_2[4M_2\gamma^2_{2}+M_1N^2]}\left\{2M_1M_2\left[-4\ell(\ell+1)\gamma_{2}+(C\gamma_{1}+2M_{1})(N^2+4\frac{M_2}{M_1}\gamma^2_{2})\right] \nonumber\right.\\+\left.
N\sqrt{M_1M_2\,}\sqrt{16M_{1}^{2}M_{2}(4M_{2}\gamma^{2}_{2}+M_{1}N^2)-C^2[-4M_{1}\ell(\ell+1)+4M_{2}\gamma^{2}_{2}+M_{1}N^2]^2\,}\right\}\,,\nonumber\\
\end{eqnarray}
where negative values has to be taken for the binding energy and of small value satisfying the inequality $E \ll M_1$ [36]. The parameters in the last equation are as follows
\begin{eqnarray}
N=1+2n+A_{I}\,\,; \gamma_{1}=A+B\,\,; \gamma_{2}=A-B\,.
\end{eqnarray}

Now we write the radial wave functions for the Hellmann potential by using Eq. (14) as
\begin{eqnarray}
R_{n\ell}\,(r)=A_{n\ell}z^{\alpha}(1-z)^{\beta}\,_{2}F_{1}(-n, n+2\alpha+2\beta+1; 1+2\alpha; z)\,,
\end{eqnarray}
with a normalization constant $A_{n\ell}$.

We want to give approximate analytical results for the bound state solutions for some special parameter values in the Hellmann potential.

\subsection{R\lowercase{esults} and D\lowercase{iscussion}}

If we set the potential parameters in Eq. (6) as
\begin{eqnarray}
A=0\,\,;B<0\,,
\end{eqnarray}
we get Yukawa-type potential having the form
\begin{eqnarray}
V(r)=B\,\frac{e^{-\,Cr}}{r}\,.
\end{eqnarray}
The parameters given in Eq. (10), and the parameter $A_I$ become
\begin{eqnarray}
\alpha^2&=&-\left(B+\frac{E}{C}\right)\left[\frac{M_2}{M_1}B+\frac{2M_2}{C}\left(1+\frac{E}{2M_1}\right)\right]\,,\nonumber\\
\beta^2&=&\ell(\ell+1)-2\frac{M_{2}E}{C^2}\left(1+\frac{E}{2M_1}\right)\,,\nonumber\\
A_{I}&=&\sqrt{\left(\ell+\frac{1}{2}\right)^2-\frac{M_2}{M_1}\,B^2\,}\,,
\end{eqnarray}
and the energy expression of the spinless Salpeter equation for the Yukawa-type potential becomes
\begin{eqnarray}
E=-\frac{1}{2M_2[B^{2}M_2+M_1N^2]}\left\{BCM_{1}M_{2}\ell(\ell+1)+(BC+2M_1)\left[B^2M^{2}_{2}+M_{1}M_{2}N^2\right] \nonumber\right.\\+\left.
N\sqrt{M_1M_2\,}\sqrt{4M_{1}^{2}M_{2}(M_{2}B^{2}+M_{1}N^2)-C^2[M_{1}\ell(\ell+1)-M_{2}B^{2}+M_{1}N^2]\,}\right\}\,,\nonumber\\
\end{eqnarray}
with $N=n+\frac{1}{2}-A_{I}$.

In order to obtain the results for the bound state solutions of the Coulomb-type potential, we expand Eq. (6) into a series for $C \rightarrow 0$
\begin{eqnarray}
V(r)\sim \frac{B-A}{r}-BC+\ldots\,,
\end{eqnarray}
which is a shifted Coulomb-type potential. We restrict ourself to the case where $B < A$ for discussing the bound states. Inserting first two terms in Eq. (24) into Eq. (4) and using the definitions
\begin{eqnarray}
-\delta^2&=&2M_{2}BC\left(1+\frac{E}{M_1}\right)+\frac{M_2}{M_1}\,B^2C^2+2M_{2}E\left(1+\frac{E}{2M_1}\right)\,,\nonumber\\
\sigma&=&2\,\frac{M_2}{M_1}\,BC(B-A)+2M_{2}(B-A)\left(1+\frac{E}{M_1}\right)\,,\nonumber\\
-\eta&=&-\ell(\ell+1)+\frac{M_2}{M_1}(B-A)^2\,,
\end{eqnarray}
we get
\begin{eqnarray}
\frac{d^2R_{n\ell}\,(r)}{dr^2}-\left(\delta^2+\frac{\sigma}{r}+\frac{\eta}{r^2}\right)R_{n\ell}\,(r)=0\,.
\end{eqnarray}
Using a trial wave function $R_{n\ell}\,(r)=r^{L}e^{-\,\delta r}f(r)$, and defining a new variable as $z=2\delta r$ gives us from the last equation
\begin{eqnarray}
z\frac{d^2f(z)}{dz^2}+(2L-z)\frac{df(z)}{dz}-\left(L+\frac{\sigma}{2\delta}\right)f(z)=0\,,
\end{eqnarray}
where we set $\eta=L(L-1)$. The differential equation in Eq. (27) is a Kummer-type equation having the form [35]
\begin{eqnarray}
x\frac{d^2y(x)}{dx^2}+(c-x)\frac{dy(x)}{dx}-ay(x)=0\,.
\end{eqnarray}

The general solution of Eq. (28) is obtained by expanding in a power series around $x=0$, and we have [37]
\begin{eqnarray}
y(x)=N'\,_{1}F_{1}(a;c;x)+N''x^{1-c}\,_{1}F_{1}(a-c+1;2-c;x)\,,
\end{eqnarray}
where $\,_{1}F_{1}(a;c;x)$ is the confluent hypergeometric function having the explicit form [37]
\begin{eqnarray}
\,_{1}F_{1}(a;c;x)=1+\frac{1}{1!}\frac{a}{c}\,x+\frac{1}{2!}\frac{a(a+1)}{c(c+1)}\,x^2+\ldots=\sum_{k=0}^{\infty}\frac{(a)_{k}}{(c)_{k}k!}z^{k}\,,
\end{eqnarray}

Because of the physical requirements on the wave function in Eq. (27), \textit{i.e.}, $R_{n\ell}\,(r)$ has to be finite for $r \rightarrow 0$ and $r \rightarrow \infty$, we set $N''=0$. By comparing Eq. (27) with Eq. (28), we can write the solution of Eq. (27)
\begin{eqnarray}
f(z) \sim \,_{1}F_{1}\left(L+\frac{\sigma}{2\delta}; 2L; z\right)\,.
\end{eqnarray}

We can obtain a finite solution from Eq. (30) if we write in Eq. (31)
\begin{eqnarray}
L+\frac{\sigma}{2\delta}=-n\,\,;n=0, 1, 2, \ldots\,.
\end{eqnarray}

The last equation gives the quantization condition and, with the help of Eq. (25), gives us the energy spectrum of the spinless Salpeter equation for the shifted Coulomb-type potential as
\begin{eqnarray}
E=-\frac{1}{\Lambda(M_1,M_2,A,B)}\left[(M_1+BC)\Lambda(M_1,M_2,A,B)+M_1(n+L)\sqrt{M_{1}\Lambda(M_1,M_2,A,B)\,}\right]\,,\nonumber\\
\end{eqnarray}
with
\begin{eqnarray}
\Lambda(M_1,M_2,A,B)=M_1(n+L)^2+M_2(B-A)^2\,.
\end{eqnarray}

If the second term in Eq. (24) is neglected, then we obtain the energy eigenvalues for attractive Coulomb-type potential of the spinless Salpeter equation as
\begin{eqnarray}
E=-M_{1}\left[1+(n+L)\sqrt{\frac{M_{1}}{\Lambda(M_1,M_2,A,B)}\,}\,\right]\,,
\end{eqnarray}
which corresponds to the result obtained from Eq. (31) for $C \rightarrow 0$. These two results show the contribution to the energy coming from the exponential part of the potential clearly.

\section{Conclusions}
We have obtained the bound state solutions of the spinless Salpeter equation by using an approximation on the centrifugal term. We have written the corresponding radial wave functions in terms of the hypergeometric functions $\,_{1}F_{2}(a,b;c;z)$. We have also given the results for the Yukawa-type potential by setting the potential parameters as $A=0, B<0$. In order to have the results for the Coulomb-type potential, we have used the normalization condition obtained from the finite solution of the Kummer differential equation.

\section{Acknowledgements}
The author would like to thank Prof Dr Ramazan Sever for his valuable time and the reviewer for her/his patience and comments that improved the paper.

\end{document}